\documentclass[aps,prb,twocolumn,groupedaddress,showpacs]{revtex4-1}
\Large   
\usepackage{graphicx}
\begin{document}

\title{Superconducting Properties of BaBi$_{3}$}

\author{Neel Haldolaarachchige, S.~K.~Kushwaha, Quinn Gibson and R.~J.~Cava}
\affiliation{Department of Chemistry, Princeton University, Princeton, New Jersey 08544, USA}

\begin{abstract}  
We report the superconducting properties of single crystals of the intermetallic perovskite-related compound BaBi$_{3}$. The superconducting transition temperature ($T_{c}=5.82$~K) was obtained from heat capacity measurements. Using the measured values for the critical fields $H_{c1}, H_{c2}$, and the specific heat $C$, we estimate the thermodynamic critical field $H_{c}$(0), coherence length $\xi$(0), Debye temperature $\Theta  _{D}$ and coupling constant $\lambda  _{ep}$. $\Delta C/\gamma T_{c}$ and $\lambda _{ep}$ suggest that BaBi$_{3}$ is a moderately coupled superconductor and $\gamma $ suggests an enhanced density of states at the Fermi level. Electronic band structure calculations show a complex Fermi surface and a moderately high DOS at the Fermi level. Further analysis of the electronic specific heat shows that the superconducting properties are dominated by \textit{s}-wave gap. 
\end{abstract}

\maketitle
\newcommand{\angstrom}{\mbox{\normalfont\AA}}

\section{Introduction}
Perovskite structure compounds are among the most widely studied superconductors. Interest in non-oxide perovskites as superconductors was enhanced with the discovery of superconductivity in MgCNi$_{3}$, leading to experimental and theoretical studies on related systems.~\cite{mgcni32001, mollah2004, bakbio31988, mgb22001} 
Alkali-Bi based compounds have recently attracted attention because the characterization of the electronic structure of Na$_{3}$Bi has led to its proposal as a Dirac Semimetal.~\cite{na3bi2012, na3bi2014, na3bisatya} Superconductors based on heavy metals such as Bi are potentially interesting because their properties may be significantly influenced by spin orbit coupling.

Here we report a detailed study of the superconducting properties of BaBi$_{3}$, supported by electronic structure calculations. 
BaBi$_{3}$ is a previously reported superconductor, but it has not so far been the subject of detailed experimental or theoretical studies.~\cite{mattias1957, mattias1952} This material crystallizes in a perovskite-related crystal structure~\cite{babi31961, babi31970} which consists of an array of corner sharing the Bi$_{6}$ octahedra, with no atoms inside, and Ba in the perovskite A site; the structure is related to that of both the BaBiO$_{3}$ oxide perovskite~\cite{bakbio31988} and MgCNi$_{3}$ intermetallic perovskite~\cite{mgcni32001} superconductors, but with the octahedral sites vacant.
The Bi$_{6}$ octahedra can be considered as creating a three dimensional framework. Since the dominant element is bismuth, a heavy element, our interest is to study this system as a possible candidate for strongly coupled heavy fermionic behavior. 

\section{Experiment and Calculation}
BaBi$_{3}$ single crystals were prepared by a Bi-self-flux method starting from elemental Ba (99.99$\%$; Alfa Aesar) and Bi (99.99$\%$; Alfa Aesar) pieces.
The starting materials (Ba:Bi ratio 1:4) were added into carbon-coated quartz tube inside an Ar-filled glove box. The sealed tubes were slowly heated at 600~$^{0}$C for 6 hrs. They were then slowly cooled to 330~$^{0}$C over a period of 5 days. Finally, the excess Bi-flux was removed by decanting. 
Cubic shape single crystals (1 mm$^{3}$) were observed, and were kept inside the glove box until characterization. Such handling is necessary to avoid decomposition. The purity and cell parameters of the samples were evaluated by powder X-ray diffraction (PXRD) at room temperature on a Bruker D8 FOCUS diffractometer (Cu$~K_{\alpha})$. 

The electrical resistivity was measured using a standard four-probe dc technique with an excitation current of 10 mA; small diameter Pt wires were attached to the sample using a conductive epoxy (Epotek H20E). Data were collected from 300 - 2 K and in magnetic fields up to 3 T using a Quantum Design Physical Property Measurement System (PPMS). The specific heat was measured between 0.4 and 20 K in the PPMS equipped with a $^{3}$He cryostat, using a time-relaxation method, at 0 and 5 Tesla applied magnetic fields. Magnetic susceptibility was measured in a constant magnetic
field of 20 Oe; the sample was cooled down to 2 K in zero-field, and then magnetic field was applied, followed by heating to 8 K [zero-field-cooled (ZFC)] and then cooled down again to 2 K [field-cooled (FC)] in the PPMS.
The electronic structure calculations were performed by density functional theory (DFT) using the WIEN2K code with a full-potential linearized augmented plane-wave and local orbitals [FP-LAPW + lo] basis~\cite{blaha2001,sign1996,madsen2001,sjosted2000} together with the PBE parameterization~\cite{perdew1996} of the GGA, including spin orbit coupling (SOC). The plane-wave cutoff parameter R$_{MT}$K$_{MAX}$ was set to 8 and the Brillouin zone was sampled by 20,000 k points. Convergence was confirmed by increasing both R$_{MT}$K$_{MAX}$ and the number of k points until no change in the total energy was observed. The Fermi surface was plotted with the program Crysden.

\section{Results and Discussion}
\begin{figure}[t]
  \centerline{\includegraphics[width=0.5\textwidth]{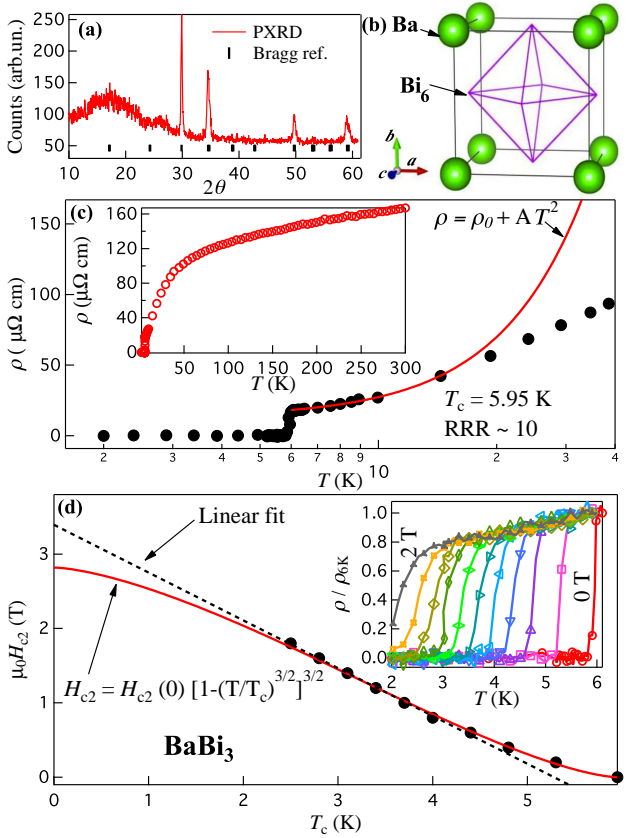}}
  \caption
    {
      (Color online) (a) PXRD pattern of BaBi$_{3}$ single crystals. (b) 3D crystal structure of BaBi$_{3}$. (c) Semi log plot of resistivity as a function of temperature for BaBi$_{3}$. The solid line represents the Fermi-liquid fit, with $\rho =\rho _{0}+AT^{n}$. (d) Magnetoresistance analysis of BaBi$_{3}$ single crystals. The main panel shows $\mu _{0}H_{c2}$ as a function of $T_{c}$ and the inset shows resistivity as a function of temperature with applied magnetic fields up to 2 Tesla. 
    }
  \label{Fig1}
\end{figure}

Fig.~\ref{Fig1} shows the resistivity, PXRD analysis of BaBi$_{3}$ single crystals and the 3D crystal structure. BaBi$_3$ has tetragonal symmetry. (P4/$mmm$, space group number 123) Fig.~\ref{Fig1}(a) shows that the PXRD pattern of the crystals employed (crushed for the PXRD pattern) matches the standard pattern in the ICSD database (code number 58634).~\cite{babi31961} (The “hump” in the low 2$\theta$ range of the PXRD pattern is due to the paratone-oil that covers the sample to prevent it from decomposing during the pattern acquisition.) A schematic view of crystal structure of BaBi$_{3}$ is shown in Fig.~\ref{Fig1}(b). The corner sharing Bi$_{6}$-octahedra that make the 3D-bismuth network are easily discerned.

\begin{figure}[t]
  \centerline{\includegraphics[width=0.5\textwidth]{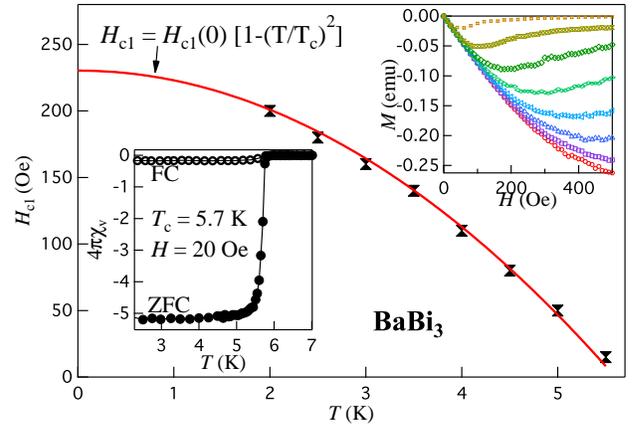}}
  \caption
    {
      (Color online) Analysis of the temperature dependent DC-magnetization of a single crystal of BaBi$_{3}$. The main panel shows $\mu _{0}H_{c1}$ as a function of $T_{c}$. The upper right inset shows the DC-magnetization as a function of applied magnetic field at different temperatures below the superconducting T$_{c}$. The lower left inset shows the ZFC and FC data through T$_{c}$.
    }
  \label{Fig2}
\end{figure}

Fig.~\ref{Fig1}(c) shows the temperature dependent resistivity from 300 K to 2 K. Metallic behavior $\left( \frac{d\rho}{dT} > 0\right)$ can be observed in the normal state resistance of the BaBi$_{3}$ single crystals. An \textit{S}-like inflection point can be observed around 40 K, with room temperature $\rho = 160~\mu\Omega~cm$ and residual resistivity ratio $\left( RRR=\frac{R_{300 K}}{R_{6 K}} = 10\right) $. Similar behavior is observed in other Bi-based superconductors~\cite{rhbi2012} and the perovskite-type intermetallic system MgCNi$_{3}$.~\cite{mgcni32001}
The $\rho (T)$ (inset in Fig.~\ref{Fig1}(c)) shows a tendency to saturate at high temperature with convex curve above 50 K. The behavior could be related to the Ioffe-Regel
limit,~\cite{ioffe} when the charge carrier mean free path is comparable to the interatomic spacing and/or to the two-band conductivity.~\cite{zverev} The low-temperature resistivity data can be described by the power law $\rho = \rho _{0} + AT^{n}$ with $n=2$, the residual resistivity $\rho _{0} = 12~\mu \Omega ~cm$, and the coefficient $A = 0.014~\frac{\mu\Omega cm}{K^{2}}$. The Fermi-liquid fit is shown as
solid lines in Fig.~\ref{Fig1}(a).
The value of \textbf{A} is often taken as a measure of the degree of electron correlations. The value found here suggests that BaBi$_{3}$ is a weakly correlated electron system; the variation of the low temperature resistivity as $T^{2}$ indicates Fermi-liquid behavior.~\cite{kittel}

Fig.~\ref{Fig1}(d) shows an analysis of the magnetoresistance data for a BaBi$_{3}$ single crystal. The width of the superconducting transition increases slightly with increasing magnetic field. Selecting the 50$\%$ normal state resistivity drop point as the transition temperature, we estimate the orbital upper critical field, $\mu _{0}H_{c2}$(0), from the Werthamer-Helfand-Hohenberg (WHH) expression, 
$\mu _{0}H_{c2}(0)=-0.693~T_{c}\frac{dH_{c2}}{dT}\vert_{T=T_{c}}$. A nearly linear relationship is observed in Fig.~\ref{Fig1}(d) between $\mu _{0}H_{c2}$ and $T_{c}$. The slope is used to calculate $\mu _{0}H_{c2}(0)=$~2.82~T. The value of $\mu _{0}H_{c2}(0)$ is smaller than the weak coupling Pauli paramagnetic limit $\mu _{0}H^{Pauli} = 1.82~T_{c} = 10.856$ T for BaBi$_{3}$. 
We also used the empirical formula $H_{c2}(T)=H_{c2}(0)\left[1-\left(\frac{T}{T_{c}}\right)^{\frac{3}{2}}\right]^{\frac{3}{2}}$
to calculate orbital upper critical field $\left( \mu _{0}H_{c2}(0)=2.91 T\right) $, which yields a value that agrees well with the calculated value using the WHH method. These models are widely used to calculate the $\mu _{0}H_{c2}(0)$ for a variety of intermetallics and oxide superconductors.~\cite{amar2011, amar2010, lan2001, neel2014, maz2013, daigo2013, huxia2013} 
Also, for a one-band superconductor, the orbital upper critical field derived from the slope $\left( \textit{k}=\frac{dH_{c2}}{dT}\mid_{T=T_{c}}\right) $ of the \textit{H-T} phase boundary at \textit{T}$_{c}$ is an indication of clean limit $\left( \frac{\mu_{0}H_{c2}}{k~T_{c}}=-0.73\right) $or dirty limit $\left( \frac{\mu_{0}H_{c2}}{k~T_{c}}=-0.69\right) $ behavior. BaBi$_{3}$ has the $\left( \frac{\mu_{0}H_{c2}}{k~T_{c}}=-0.74\right) $value, therefore BaBi$_{3}$ is a type II BCS-superconductor that is close to the clean limit.~\cite{cdtheory1966, cdmgb22002} 
$\left( \frac{\mu_{0}H_{c2}}{k~T_{c}}\right)$ is not very sensitive to the coupling strength, and this result is actually close to that of strong-coupling superconductors.~\cite{carbotte, cdmgb22002} 
The upper critical field value $\mu _{0}H_{c2}(0)$ of the BaBi$_{3}$ superconductor can be used to estimate the Ginzburg-Landau coherence length $\xi (0)=\sqrt{\Phi _{0}/2\pi H_{c2}(0)}=110.45$~\AA, where $\Phi _{0}=\frac{hc}{2e}$ is the magnetic flux quantum.~\cite{clogston1962,werthamer1966} This value is comparable to that for the alkali-Bi superconductor-NaBi and larger than that of MgCNi$_{3}$ (see Table.~\ref{tab:1}).~\cite{mgcni32001, satya2014}

Fig.~\ref{Fig2} shows the analysis of the DC-magnetization of BaBi$_{3}$ single crystals. The bulk superconducting transition T$^{onset} _{c}$ = 5.8 K can clearly be seen in the lower left inset of the Fig.~\ref{Fig2}. Very similar values of T$_{c}$ from both resistivity and susceptibility confirm that our single crystals are very high quality. A high superconducting shielding fraction can be observed with zero-field-cooled (ZFC-shielding) and field-cooled (FC-Meissner) data in the figure. The susceptibility in the normal state is Pauli paramagnetic like, with a small moment ($ \chi = 0.003 \frac{cm^{3}}{mol Oe}$).
The magnetization as a function of magnetic field over range of temperature below the superconducting T$_{c}$ is shown in the upper right inset of Fig.~\ref{Fig2}. For analysis of the lower critical field the point of 2.5\% deviation from the full shielding effect was selected for each temperature. The main panel of the Fig.~\ref{Fig2} shows $\mu_{0}H_{c1}$ as a function of T$_{c}$. The lower critical field behavior was analyzed with the equation $H_{c1}(T)=H_{c1}(0)\left[1-\left(\frac{T}{T_{c}}\right)^{2}\right]$. The $\mu _{0}H_{c1}$ data is well described with the above equation and a least square fit yielded the value of $\mu _{0}H_{c1}(0)$=230 Oe, which is larger than that of the both MgCNi$_{3}$ and NaBi superconductors (see Table.~\ref{tab:1}).  

\begin{figure}[t]
  \centerline{\includegraphics[width=0.5\textwidth]{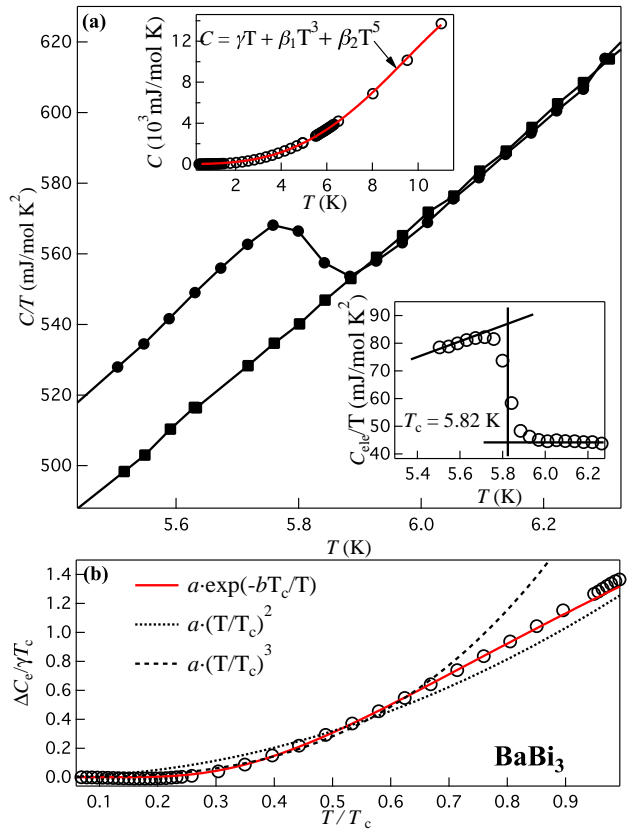}}
  \caption
    {
      (Color online) Analysis of the heat capacity of a BaBi$_{3}$ single crystal. (a) The main panel shows the heat capacity in 0 T and 5 T fields. The upper left inset shows the heat capacity data fit with the equation $C=\gamma T+\beta_{1} T^{3}+\beta_{2} T^{5}$. The lower right inset shows the heat capacity jump at T$_{c}$. (b) The heat capacity data below T$_{c}$, down to 0.4 K. Analysis suggests that BCS type \textit{s}-wave pairing symmetry dominates the superconducting state.
    }
  \label{Fig3}
\end{figure}

\begin{figure}[t]
  \centerline{\includegraphics[width=0.5\textwidth]{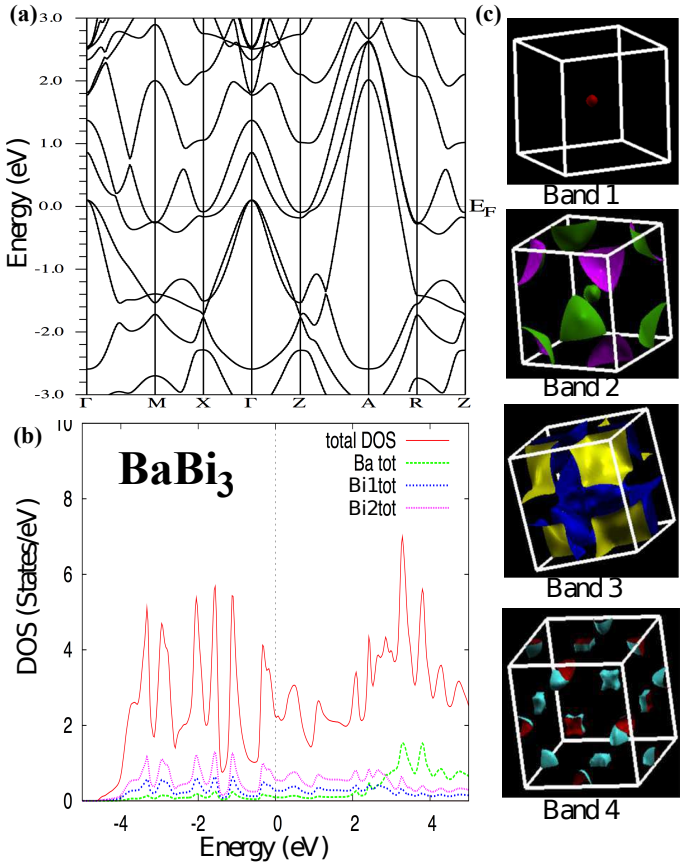}}
  \caption
    {
      (Color online) (a) Band structure of BaBi$_{3}$. (b) Total and partial DOS analysis of BaBi$_{3}$. (c) The Fermi surfaces of the four bands through the Fermi level. 
    }
  \label{Fig4}
\end{figure}

Fig.~\ref{Fig3} shows the characterization of the superconducting transition by specific heat measurements. Fig.~\ref{Fig3}(a) shows $\frac {C}{T}$ as a function of $T$, characterizing the specific heat jump at the thermodynamic transition. This jump is completely suppressed under a 5 T applied magnetic field. The superconducting transition temperature $T_{c}$ = 5.8 K is shown in the lower left inset of Fig.~\ref{Fig3}(a), as extracted by the standard equal area construction method. 
The low temperature normal state specific heat is non-Debye-like. Non-Debye behavior has often been reported in superconductors and is ascribed either to a large Einstein contribution or a low Debye temperature $\theta _{D}$. Because an Einstein phonon contribution is negligible below 20 K, a second term in the harmonic-lattice approximation is needed in cases such as ours to improve the Debye phonon specific heat.~\cite{lifeas2012, nibi32000, srsn42011, pdbi22012}
We find that the low temperature normal state specific heat can be well fitted with $\frac{C}{T} = \gamma _{n} + \beta_{1} T^{2} + \beta_{2} T^{4}$,
where $\gamma _{n} T$ represents the electronic contribution in normal state and $\beta_{1} T^{3}$ and $\beta_{2} T^{5}$ describe the lattice-phonon contributions to the specific heat. 
The solid line in Fig.~\ref{Fig3}(a) shows the fitting; the electronic specific heat coefficient $\gamma _{n} = 41 \frac{mJ}{mol~K^{2}}$ and the phonon/lattice contributions $\beta_{1} = 16 \frac{mJ}{mol~K^{4}}$ and $\beta_{2} = -0.055 \frac{mJ}{mol~K^{4}}$ are extracted from the fit. 
The value of $\gamma$ obtained is relatively larger than that of MgCNi$_{3}$ (see Table.~\ref{tab:1}).~\cite{mgcni32001} 
The high value of $\gamma$ is an indication of a moderately enhanced density of states near the Fermi level, which is supported by our band structure calculations (see below). 

The ratio $\frac{\Delta C}{\gamma T_{c}}$ can be used to measure the strength of the electron-phonon coupling.~\cite{padamsee1973} The specific heat jump $\frac{\Delta C}{T_{c}}$ for the sample is about 35 $\frac{mJ}{mol~K^{2}}$, setting the value of $\frac{\Delta C}{\gamma~T_{c}}$ to 0.86. This is smaller than the weak-coupling limit of 1.43 for a conventional BCS superconductor and is comparable to that of MgCNi$_{3}$ (see Table.~\ref{tab:1}). The results suggest that BaBi$_{3}$ is a moderately electron$-$phonon coupled superconductor.
The relatively small value of the heat capacity jump at T$_{c}$ indicates that the quasiparticles participating in the SC condensation experience strong elastic scattering, because inelastic scattering usually enhances the specific heat jump ratio; similar behavior observed in some other superconducting systems.~\cite{lifeas2012}

In a simple Debye model for the phonon contribution to the specific heat, the $\beta$ coefficient is related to the Debye temperature $\Theta _{D}$ through $\beta = nN_{A}\frac{12}{5}\pi ^{4}R\Theta _{D}^{-3}$, where $R = 8.314~\frac{J}{mol~K}$, $\textit{n}$ is the number of atoms per formula unit and $N_{A}$ is Avogadro\textquoteright s number. The calculated Debye temperature for BaBi$_{3}$ is thus 142 K. This value of the Debye temperature is comparable to that of elemental Bi and Bi-based superconductors such as NaBi and NiBi$_{3}$, but is slightly smaller than that of the MgCNi$_{3}$.~\cite{satya2014, bi1966, nibi32000, mollah2004} 
An estimation of the strength of the electron-phonon coupling can be derived from the McMillan formula
$\lambda _{ep} = \frac{1.04 + \mu ^{*} ln\frac{\Theta _{D}}{1.45T_{c}}}{(1-0.62\mu ^{*}) ln\frac{\Theta _{D}}{1.45T_{c}}-1.04}$.~\cite{mcmillan1968,poole1999} 
McMillan\textquoteright s model contains the dimensionless electron-phonon coupling constant $\lambda _{ep}$, which, in the Eliashberg theory, is related to the phonon spectrum and the density of states. This parameter $\lambda _{ep}$ represents the attractive interaction, while the second parameter $\mu ^{*}$ accounts for the screened Coulomb repulsion.
Using the Debye temperature $\Theta _{D}$, critical temperature $T_{c}$, and making the common assumption that $\mu ^{*} = 0.15$,~\cite{mcmillan1968} the electron-phonon coupling constant ($\lambda _{ep}$) obtained for BaBi$_{3}$ is 0.83, which suggests moderately enhanced electron-phonon coupling behavior and agrees well with $\frac{\Delta C}{\gamma T_{c}} = 0.85$.

The electronic specific heat ($C_{e}$) is further analyzed to study the superconducting gap function. $\Delta C_{e}$ was fitted to the forms $e^{-b/T}$, T$^{2}$, and T$^{3}$, which are the expected temperature dependencies for gaps that are isotropic, contain line nodes, or linear point nodes, respectively. The data agrees well with the exponential fit $\frac{\Delta C_{e}}{\gamma T_{c}}$ below T/T$_{c}$ = 0.8, while the other two fits do not describe the data. 
C$_{e}$(T) can also be well described with the semiempirical approximation, the so-called $\alpha$ model, C$_{e}$(T) = A \textit{exp}($\frac{-\Delta (0)}{k_{B}T}$), where \textit{k}$_{B}$ is the Boltzmann constant and $\Delta (0)$ is the superconducting gap at 0 K.~\cite{padamsee1973} This equation allows for varying the coupling strength $\alpha = \frac{\Delta (0)}{k_{B}T_{c}}$, instead of fixing it at the BCS weak-coupling limit, $\alpha$ = 1.76. The obtained coupling strength $\alpha$ = 1.46 (2$\Delta (0)$=2.93 k$_{B}$T$_{c}$) gives an excellent fit, which agrees well with the BCS-weak coupling limit.
Therefore, the above analysis suggests that BaBi$_{3}$ is a BCS-type isotropic-gapped superconductor. This further confirms that the superconducting properties of this system are dominated by \textit{s}-wave pairing symmetry.~\cite{tomasz2007, amar2011, daigo2013} 

Fig.~\ref{Fig4} shows the analysis of the electronic density of states for BaBi$_{3}$ based on the electronic structure calculations. Fig.~\ref{Fig4}(a) shows the band structure in the vicinity of Fermi energy E$_{F}$. According to the calculated band structure, BaBi$_{3}$ is a three-dimensional metal; 4 bands with large dispersion cross the Fermi level. The bands at the Fermi level are all derived from Bi \textit{p}-orbitals. 
The total DOS (see Fig.~\ref{Fig4}(b)) shows that the Fermi level is located near the edge of a local maximum. The value of the DOS at E$_{F}$ is in the range of 2.1-4.0
eV/f.u. (f.u. = formula unit), which agrees well with the enhanced $\gamma $ value observed from the heat capacity data. 
The partial DOS shows that the total DOS is dominated by the contributions from the two types of Bi atoms in the tetragonal symmetry structure and that the contribution from the Ba atom near the Fermi level is almost negligible. 
Fig.~\ref{Fig4}(c) shows the calculated Fermi surfaces of the four bands that crosses the Fermi level in BaBi$_{3}$. This suggests that the combination of all four bands creates a very complex Fermi surface in this compound.

\begin{table}[t]
\caption{Superconducting Parameters of perovskite-BaBi$_{3}$. Comparison is done with the perovskite-MgCNi$_{3}$ and cubic-NaBi. Superconducting parameters of MgCNi$_{3}$ and NaBi were extracted from Refs. 03 and 07 respectively }
  \centering  
  \begin{tabular}{ lc   c   c   c   c }
  \hline \hline 
    Parameter & Units & BaBi$_{3}$ & MgCNi$_{3}$ & NaBi \\  
      &   &  &  Ref.03  &  Ref.07  \\ \hline  \hline  
    $T_{c}$ & K & 5.9  &  6.7-8  &  2.1  \\
    $\rho _{0}$ & $\mu\Omega cm$ &  13  &   &    \\
    $\frac{dH_{c2}}{dT}\vert _{T=T_{c}}$ & $T~K^{-1}$ & -0.67  &    \\
    $\mu _{0}H_{c1}(0)$ & Oe & 230  & 10-12.6  &  30  \\
    $\mu _{0}H_{c2}(0)$ & T & 2.82  &  11.6-16  &  0.025   \\
    $\mu _{0}H^{Pauli}$ & T & 10.85  & 12-14.4  &  3.78  \\
    $\mu _{0}H (0)$ & T & 0.15  & 0.19  &   \\
    $\xi (0)$ & \AA & 110.4  & 45-56  &    \\
    $\lambda (0)$ & \AA & 1340  & 1800-2480  &     \\
    $\kappa (0)$ & \AA & 12.1  & 43-66  &   \\
    $\gamma (0)$ & $\frac{mJ}{mol~K^{2}}$ & 41  &  10  &  3.4   \\
    $\frac{\Delta C}{\gamma T_{c}}$ & & 0.85  &  1.9  &  0.78  \\
    $\Theta _{D}$ & K & 142  &  284  &  140   \\
    $\lambda _{ep}$ &  & 0.82  & 0.66-0.84  &  0.62  \\
     $\frac{2 \Delta (0)}{k_{B}T_{c}}$ &  & 2.93  &   &    \\  
    $N(E_{F})$ & $\frac{eV}{f.u.}$ & 1.8  &     &  0.88   \\ \hline \hline
  \end{tabular}
  \label{tab:1}
\end{table}

The superconducting parameters are presented in Table.~\ref{tab:1}. Comparison with MgCNi$_{3}$ is given because it shares perovskite~-~type structure with BaBi$_{3}$ and comparison to NaBi is given because both compounds are alkaline-bismuth based superconducting compounds. 
The superconducting parameters of BaBi$_{3}$ are more close to the perovskite-MgCNi$_{3}$ superconducting compound. It is interesting to see that electron-phonon coupling constants of BaBi$_{3}$ and MgCNi$_{3}$ are very slimilar. 
Susceptibility at 10 K ($\chi = 1.03 \times 10^{-4} \frac{m^{3}}{mol}$) can be considered as the spin susceptibility. This allows for an estimate of the Wilson ratio $R_{W} = \frac{\pi ^{2} k_{B}^{2} \chi _{spin}}{3\mu _{B}^{2}\gamma }$, which is smaller than that of the free electron value of 1. Also, the coefficient of the quadratic resistivity term can be normalized with the effective mass term from the heat capacity, which gives the Kadowaki-Woods ratio $\frac{A}{\gamma ^{2}}$. This ratio is found to be 0.5~$a_{0}$, where $a_{0}$ = 10$^{-5} \frac{\mu \Omega ~cm}{(mJ / mol~K)^{2}}$. $R_{W}$ and $\frac{A}{\gamma ^{2}}$ both indicate that BaBi$_{3}$ is a weakly-correlated electron system.~\cite{jacko2009,wilson1975,yamada1975}
The value of $\gamma$ extracted from the measured specific heat data corresponds to an electronic density of states at the Fermi energy $N(E_{F})(exp)$ of 1.8 states/(eV f.u.), as estimated from the relation~\cite{kittel, poole1999} $\gamma = \pi ^{\frac{3}{2}} k_{B}^{2} N(E_{F}) (1+\lambda _{ep})$. This value is comparable to the value obtained from our band structure calculation $N(E_{F})(cal)$ of 2.2 states/(eV f.u.). 
A straightforward calculation of the condensation energy from the relation $\textit{U}(0) = \frac{1}{2}\Delta ^{2}(0)N(E_{F})$ produces a value of $\textit{U}(0)=1190\frac{mJ}{mol}$.~\cite{amar2010, tinkham}
Using the upper and lower critical fields and the relation $H_{c2}(0)H_{c1}(0) = H_{c}(0)^{2}[ln \kappa (0) + 0.08]$,~\cite{tinkham}  the thermodynamic critical field H$_{c}$(0) was found to be 0.15 T. This value is comparable to that of MgCNi$_{3}$.
Assuming \textit{g}=2 for conduction electrons, one can estimate the Pauli limiting field for BaBi$_{3}$ from the relation of $\mu_{0}H^{Pauli} = \frac{\Delta (0)}{\mu_{B}~1.41}$ = 11.7 T. This is very close to the value obtained from the orbital upper critical field. 
The actual H$_{c2}$ of real materials is generally influenced by the both orbital and spin-paramagnetic
effects. The relative importance of the orbital and spin-paramagnetic effects can be described by the
Maki parameter,~\cite{maki}
which can be calculated by $\frac{\mu_{0}H_{c2}}{H^{Pauli}}=\frac{\alpha}{1.41}=0.3$.~\cite{carbotte, cdmgb22002, junod} Even though BaBi$_{3}$ shows enhanced density of states at the Fermi level $\alpha <  1$ indicates that this system dose not represent heavy electron mass or multiple small Fermi pockets, which is also consistent with our calculation.  


\section{Conclusion}
We have grown single crystals and characterized the superconducting properties of perovskite-like BaBi$_{3}$. A bulk superconducting transition was confirmed and characterized through magnetization and heat capacity measurements on the single crystals. The electronic contribution to the specific heat is relatively large, $\gamma$ = 41 $\frac{mJ}{mol~K^{2}}$ and the electron-phonon coupling constant is $\lambda _{ep}=0.86$. BCS-type \textit{s}-wave paring symmetry is inferred from the behavior of the electronic heat capacity in the superconducting state. The electronic density of states calculation suggests that BaBi$_{3}$ is a good 3D-metal and the Bi \textit{p}-bands are dominant through the Fermi level. Many bands can be found at energies through the Fermi level, which creates a complex Fermi surface.   

\section{acknowledgments}
This work was supported by the Department of Energy Division of Basic Energy Sciences, grant DE-FG02-98ER45706.

\end{document}